\documentclass[showpacs,amsmath,amssymb,prb,aps,twocolumn,superscriptaddress]{revtex4}
\usepackage{graphicx}
\begin{document}
\title {Inhomogeneous Gilbert damping from impurities and electron-electron interactions}
\author {E. M. Hankiewicz}
\email{hankiewicz@fordham.edu}
\affiliation{Department of Physics, Fordham University, Bronx, New York 10458, USA}
\affiliation{Department of Physics and Astronomy, University of Missouri, Columbia, Missouri 65211, USA}
\author{G. Vignale}
\affiliation{Department of Physics and Astronomy, University of Missouri, Columbia, Missouri 65211, USA}
\author{Y. Tserkovnyak}
\affiliation{Department of Physics and Astronomy, University of California, Los Angeles, California 90095, USA}
\date{\today}

\begin{abstract}
We present a unified theory of magnetic damping in itinerant
electron ferromagnets at order $q^2$ including  electron-electron
interactions and disorder scattering. We show that the Gilbert
damping coefficient can be expressed in terms of the spin
conductivity, leading to a Matthiessen-type formula in which
disorder and interaction contributions are additive. In a weak
ferromagnet regime, electron-electron interactions lead to a
strong enhancement of the Gilbert damping.
\end{abstract}


\maketitle

{\em Introduction} -- 
In spite of much effort, a complete theoretical description of the
damping of ferromagnetic spin waves in itinerant electron
ferromagnets is not yet available.~\cite{Yaroslav05} Recent
measurements of  the dispersion and damping of spin-wave
excitations driven by a direct spin-polarized current  prove that
the theoretical picture is incomplete, particularly when it comes
to calculating the linewidth
 of these excitations.\cite{Krivorotov07}  One of the most important parameters of the theory is
 the so-called Gilbert damping  parameter
$\alpha$,\cite{Gilbert04}  which controls the damping rate
 and thermal noise and is often assumed to be independent of the wave vector of the excitations.
 This assumption is justified for excitations of very long wavelength (e.g., a homogeneous precession of the magnetization),
  where $\alpha$ can originate in a relatively weak spin-orbit (SO) interaction
  \cite{Hankiewicz07}.
But it becomes dubious as the wave vector $q$ of the excitations grows.
 Indeed, both electron-electron (e-e) and electron-impurity interactions can cause an
  {\it inhomogeneous} magnetization to decay into spin-flipped electron-hole pairs, giving rise
  to a $q^2$ contribution to the Gilbert damping. In practice, the presence of this contribution means that the Landau-Lifshitz-Gilbert
equation contains a term proportional to $-{\bf m} \times \nabla^2
\partial_t {\bf m}$ (where ${\bf m}$ is the magnetization)  and
requires neither spin-orbit nor magnetic disorder scattering.   By
contrast, the homogeneous damping term is  of the form ${\bf m}
\times \partial_t {\bf m}$ and vanishes in the absence of SO or
magnetic disorder scattering.

The influence of disorder on the linewidth of spin waves in
itinerant electron ferromagnets was discussed in
Refs.~\onlinecite{Singh89,Tesanovic89,Safonov}, and the role of
e-e interactions in spin-wave damping was studied in
Refs.~\onlinecite{Silin57,Mineev04} for spin-polarized liquid
He$^3$ and in Refs.~\onlinecite{Takahashi99,Qian02} for two- and
three-dimensional electron liquids, respectively. In this paper,
we present a unified semiphenomenological approach, which enables
us to calculate on equal footing the contributions of disorder and
e-e interactions to the Gilbert damping parameter to order $q^2$.
The main idea is to apply to the transverse spin fluctuations of a
ferromagnet the method first introduced by Mermin~\cite{Mermin65}
for treating the effect of disorder on the dynamics of charge
density fluctuations in metals.~\cite{Thebook}  Following this
approach, we will show that the $q^2$ contribution to the damping
in itinerant electron ferromagnets can be expressed in terms of
the transverse spin conductivity, which in turn separates into a
sum of disorder and e-e terms.

A major technical advantage of this approach is that the ladder
vertex corrections to the transverse spin-conductivity vanish in
the absence of SO interactions, making the diagrammatic
calculation of this quantity a straightforward task. Thus we are
able to provide explicit analytic expressions for the disorder and
interaction contribution to the $q^2$ Gilbert damping to the
lowest order in the strength of the interactions. Our paper
connects and unifies different approaches and gives a rather
complete and simple theory of $q^2$ damping. In particular, we
find that for weak metallic ferromagnets the $q^2$ damping can be
strongly enhanced by e-e interactions, resulting in a value
comparable to or larger than typical in the case of homogeneous
damping. Therefore, we believe that the inclusion of a damping
term proportional to $q^2$ in the phenomenological Landau-Lifshitz
equation of motion for the magnetization\cite{landauBOOKv9} is a
potentially important modification of the theory in strongly
inhomogeneous situations, such as current-driven
nanomagnets\cite{Krivorotov07} and the ferromagnetic domain-wall
motion\cite{tserkovJMMM08}. \footnote{In ferromagnets whose
nonuniformities are beyond the linearized spin waves, there is a
nonlinear $q^2$ contribution to damping, (see J. Foros and A.
Brataas and Y. Tserkovnyak, and G. E. W. Bauer, arXiv:0803.2175)
which has a different physical origin, related to the longitudinal
spin-current fluctuations.}

{\em Phenomenological approach} --
In Ref.~\onlinecite{Mermin65},
Mermin constructed the density-density response function of an
electron gas in the presence of impurities through the use of a
local drift-diffusion equation, whereby the gradient of the
external potential  is cancelled, in equilibrium, by an opposite
gradient of the local chemical potential. In diagrammatic
language, the effect of the local chemical potential  corresponds
to the inclusion of the vertex correction in the calculation of
the density-density response function.  Here, we use a similar
approach to obtain the transverse spin susceptibility of an
itinerant electron ferromagnet, modeled as an electron gas whose
equilibrium magnetization is along the $z$ axis.

Before proceeding we need to clarify a delicate point. The
homogeneous electron gas is not spontaneously ferromagnetic at the
densities that are relevant for ordinary magnetic
systems.~\cite{Thebook}
 In order to produce the desired equilibrium magnetization, we must therefore impose
 a static fictitious field  $B_0$.  Physically, $B_0$ is the  ``exchange" field $B_{ex}$
 plus any external/applied magnetic field $B_0^{\rm app}$ which may be additionally present.
  Therefore, in order to calculate the transverse spin susceptibility we must take into account
  the fact that the exchange field associated with a uniform magnetization
  is parallel to the magnetization and
  changes direction when the latter does.
  As a result, the actual susceptibility $\chi_{ab}(\mathbf{q},\omega)$
   differs from the susceptibility calculated at constant $B_0$,
    which we denote by $\tilde \chi_{ab}(\mathbf{q},\omega)$, according to the well-known relation:\cite{Qian02}
\begin{equation} \label{chiproper}
\chi^{-1}_{ab}(\mathbf{q},\omega) = \tilde
\chi^{-1}_{ab}(\mathbf{q},\omega) -\frac{\omega_{ex}}{M_0}
\delta_{ab}\,.
\end{equation}
Here, $M_0$ is the equilibrium magnetization (assumed to point
along the $z$ axis) and $\omega_{ex} = \gamma B_{ex}$ (where
$\gamma$ is the gyromagnetic ratio)
 is the precession frequency associated with the exchange field. $\delta_{ab}$ is the Kronecker delta.
 The indices $a$ and $b$ denote directions ($x$ or $y$) perpendicular
  to the equilibrium magnetization and $\mathbf{q}$ and $\omega$
  are the wave vector and the frequency of the external perturbation.
Here we focus solely on the calculation of the response function
$\tilde \chi$ because term $\omega_{ex}\delta_{ab}/M_0$ does not
contribute to Gilbert damping. We do not include the effects of
exchange and external fields on the orbital motion of the
electrons.

The generalized continuity equation for the
 Fourier component of the transverse spin density $M_a$ in the direction $a$ ($x$ or $y$)
 at wave vector $\mathbf q$ and frequency $\omega$ is
\begin{align}\label{continuity}
-i\omega M_{a}(\mathbf{q},\omega)=&-i\gamma\mathbf q \cdot \mathbf
j_{a}(\mathbf{q},\omega)  -\omega_0\epsilon_{ab}
M_{b}(\mathbf{q},\omega)\nonumber\\ &+ \gamma M_0\epsilon_{ab}
B^{\rm app}_{b}(\mathbf{q},\omega)\,,
\end{align}
where $B_{a}^{\rm app}(\mathbf{q},\omega)$ is the transverse
external magnetic field driving the magnetization and $\omega_0$
is the precessional frequency associated with a static magnetic
field $B_0$ (including exchange contribution) in the $z$
direction. $\mathbf j_{a}$ is the $a$th component of the
transverse spin-current density tensor and we put $\hbar=1$
throughout. The transverse Levi-Civita tensor $\epsilon_{ab}$ has
components $\epsilon_{xx}=\epsilon_{yy}=0$,
$\epsilon_{xy}=-\epsilon_{yx}=1$, and the summation over repeated
indices is always implied.

The transverse spin current is proportional to the gradient of the
effective magnetic field, which plays the role analogous to the
electrochemical potential, and the equation that expresses this
proportionality is the analogue of the drift-diffusion equation of
the ordinary charge transport theory:
\begin{equation}\label{DD_equation}
\mathbf j_{a}(\mathbf{q},\omega) = i\mathbf q \sigma_{\perp}\left[
\gamma B^{\rm
app}_{a}(q,\omega)-\frac{M_{a}(\mathbf{q},\omega)}{\tilde
\chi_{\perp}} \right]\,,
\end{equation}
where $\sigma_{\perp}$ ($=\sigma_{xx}$ or $\sigma_{yy}$)  is the
transverse dc (i.e., $\omega=0$) spin-conductivity and $\tilde
\chi_{\perp}=M_0/\omega_0$ is the static transverse spin
susceptibility in the $q \to 0$ limit.\footnote{Although both
$\sigma_\perp$ and $\tilde \chi_\perp$ are in principle tensors in
transverse spin space, they are proportional to $\delta_{ab}$ in
axially-symmetric systems|hence we use scalar notation.} Just as
in the ordinary drift-diffusion theory, the first term on the
right-hand side of Eq.~(\ref{DD_equation}) is a ``drift current,"
and  the second is a ``diffusion current," with the two canceling
out exactly in the static limit (for $q \to 0$), due to the
relation $M_a(0,0)=\gamma\tilde \chi_\perp B^{app}_a(0,0)$.
Combining Eqs.~(\ref{continuity}) and (\ref{DD_equation}) gives
the following equation for the transverse magnetization dynamics:
\begin{align}
\left(-i \omega \delta_{ab} +
\frac{\gamma\sigma_{\perp}q^2}{\tilde \chi_\perp}\delta_{ab} +
\omega_0
\epsilon_{ab}\right)M_b=&\nonumber\\&\hspace{-3cm}\left(M_0\epsilon_{ab}+
\gamma\sigma_{\perp}q^2\delta_{ab}\right)\gamma B^{\rm app}_b\,,
\end{align}
which is most easily solved by transforming to the
circularly-polarized components $M_{\pm}=M_x\pm i M_y$, in which
the Levi-Civita tensor becomes diagonal, with eigenvalues $\pm i$.
Solving in the ``+" channel, we get
\begin{equation}
M_+ =\gamma \tilde \chi_{+-}B^{app}_{+}= \frac{M_0-i\gamma
\sigma_\perp q^2}{\omega_0 - \omega - i\gamma\sigma_\perp
q^2\omega_0/M_0}\gamma B^{app}_+\,,
\end{equation}
from which we obtain to the leading order in $\omega$ and $q^2$
\begin{equation}\label{chi+-}
\tilde \chi_{+-}(q,\omega) \simeq
\frac{M_0}{\omega_0}\left(1+\frac{\omega}{\omega_0}\right)+i\omega\frac{\gamma\sigma_{\perp}q^2}{\omega_0^2}\,.
\end{equation}
The higher-order terms in this expansion cannot be legitimately
retained within the accuracy of the present approximation. We also
disregard the $q^2$ correction to the static susceptibility, since
in making the Mermin ansatz (\ref{DD_equation}) we are omitting
the equilibrium spin currents responsible for the latter.
Eq.~(\ref{chi+-}), however, is perfectly adequate for our purpose,
since it allows us to identify the $q^2$ contribution to the
Gilbert damping:
\begin{equation}\label{Gilbert}
\alpha = \frac{\omega_0^2}{M_0}\lim_{\omega \to 0}\frac{\Im m\,\tilde
\chi_{+-}(q,\omega)}{\omega}=\frac{\gamma\sigma_{\perp}q^2}{M_0}\,.
\end{equation}
Therefore, the Gilbert damping can be calculated from the dc
transverse spin conductivity $\sigma_\perp$, which in turn can be
computed from the zero-frequency limit of the transverse
spin-current|spin-current response function:
 \begin{eqnarray}\label{Resigmaperp}
\sigma_{\perp} = - \frac{1}{m^2_\ast {\cal V}} \lim_{\omega \to 0}
\frac{\Im m \langle\langle \sum_{i=1}^N \hat S_{ia}\hat p_{ia};
 \sum_{i=1}^N \hat S_{ia}\hat p_{ia}\rangle \rangle_\omega}{\omega}\,,
\end{eqnarray}
where $\hat S_{ia}$ is the $x$ or $y$ component
 of spin operator
for the $i$th electron, $\hat p_{ia}$ is the corresponding
component of the momentum operator, $m_\ast$
is the effective electron mass, $\cal V$ is the system volume, $N$ is the total electron number,
and $\langle\langle\hat A;\hat B\rangle\rangle_\omega$ represents the retarded linear response function
for the expectation value of an observable $\hat A$ under the action of a field that couples linearly to an observable
 $\hat B$.
 Both disorder and e-e interaction contributions can be
systematically included in the calculation of the
spin-current|spin-current response function. In the absence of
spin-orbit and e-e interactions, the ladder vertex corrections to
the conductivity are absent and calculation of $\sigma_{\perp}$
reduces to the calculation of a single bubble with Green's
functions
\begin{equation}\label{Green_function}
G_{\uparrow,\downarrow}(\mathbf{p},\omega)=\frac{1}{\omega-\varepsilon_{\mathbf{p}}+\varepsilon_F
\pm \omega_0/2 +i/2\tau_{\uparrow,\downarrow}}\,,
\end{equation}
where the scattering time $\tau_s$ in general depends on the spin band index $s=\uparrow,\downarrow$.
In the Born approximation, the scattering rate is proportional to the electron density of states,
and we can write $\tau_{\uparrow,\downarrow} = \tau\nu/\nu_{\uparrow,\downarrow}$, where $\nu_s$ is the spin-$s$
density of states and $\nu=(\nu_\uparrow+\nu_\downarrow)/2$. $\tau$ parametrizes the strength of the disorder scattering.
A standard calculation then leads to the following result:
\begin{equation}
\sigma_{\perp}^{\rm dis} = \frac{\upsilon^2_{F\uparrow}
+\upsilon^2_{F\downarrow}}{6(\nu^{-1}_{\downarrow}+\nu^{-1}_{\uparrow})}\frac{1}{\omega_0^2\tau}\,.
\end{equation}

This, inserted in Eq.~(\ref{Gilbert}), gives a Gilbert damping
 parameter in full agreement with what we have also calculated from a direct diagrammatic
 evaluation of the transverse spin susceptibility, i.e., spin-density|spin-density correlation function.
From now on, we shall simplify the notation by introducing a
transverse spin relaxation time
\begin{equation}\label{tauperp}
\frac{1}{\tau^{\rm dis}_\perp}
=\frac{4(E_{F\uparrow}+E_{F\downarrow})}
{3n(\nu^{-1}_{\downarrow}+\nu^{-1}_{\uparrow})}\frac{1}{\tau},
\end{equation}
where $E_{Fs}=m_\ast\upsilon^2_{Fs}/2$ is the Fermi
energy for spin-$s$ electrons and $n$ is the total electron density. In this notation,
the dc transverse spin-conductivity takes the form
\begin{equation}
\sigma_{\perp}^{\rm dis} =
\frac{n}{4m_\ast\omega_0^2}\frac{1}{\tau^{\rm dis}_\perp}\,.
\end{equation}
{\it Electron-electron interactions} -- One of the attractive
features of the approach based on Eq.~(\ref{Resigmaperp}) is the
ease with which e-e interactions can be included. In the weak
coupling limit, the contributions of disorder and e-e interactions
to the transverse spin conductivity are simply additive. We can
see this by using twice the equation of motion for the
spin-current|spin-current response function.  This leads to an
expression for the transverse spin-conductivity
(\ref{Resigmaperp}) in terms of the low-frequency
spin-force|spin-force response function:
\begin{eqnarray}\label{force-force}
\sigma_{\perp}= -  \frac{1}{m^2_\ast \omega_0^2 {\cal V}}
\lim_{\omega\to0} \frac{\Im m \langle \langle\sum_i \hat
S_{ia}\hat F_{ia};\sum_i \hat S_{ia}\hat
F_{ia}\rangle\rangle_{\omega}}{\omega}\,.
\end{eqnarray}
Here, $\hat F_{ia} = \dot{\hat p}_{ia}$ is the time derivative
  of the momentum operator, i.e., the operator of the force on the $i$th electron.
  The total force is the sum of electron-impurity and e-e interaction forces.
  Each of them, separately, gives a contribution of order $|v_{ei}|^2$ and $|v_{ee}|^2$,
   where $v_{ei}$ and $v_{ee}$ are matrix elements of the electron-impurity and e-e interactions, respectively,
  while cross terms are of higher order, e.g., $v_{ee}|v_{ei}|^2$.
 Thus, the two interactions give additive contributions to the conductivity.
In Ref.~\cite{Amico00}, a phenomenological equation of motion was
used to find the spin current in a system with disorder and
longitudinal spin-Coulomb drag coefficient. We can use a similar
approach to obtain transverse spin currents with transverse
spin-Coulomb drag coefficient $1/\tau^{ee}_\perp$.  In the
circularly-polarized basis,
   \begin{equation}\label{j_+/-}
    i(\omega \mp \omega_0)j_{\pm} =-\frac{nE}{4m_\ast}
    +\frac{j_{\pm}}{\tau^{\rm dis}_\perp} + \frac{j_{\pm}}{\tau^{ee}_\perp}\,,
    \end{equation}
 and correspondingly the spin-conductivities are
\begin{equation}\label{sigma_circular}
\sigma_{\pm}=\frac{n}{4m_\ast}\frac{1}{-(\omega\mp\omega_0)i+
1/\tau^{\rm dis}_\perp+ 1/\tau^{ee}_\perp}\,.
\end{equation}
In the dc limit, this gives
\begin{equation}\label{pheno-conductivity2}
\sigma_\perp(0) = \frac{\sigma_+ +\sigma_-}{2}=\frac{n}{4m_\ast}
\frac{1/\tau^{\rm
dis}_\perp+1/\tau^{ee}_\perp}{\omega_0^2+\left(1/\tau^{\rm
dis}_\perp+1/\tau^{ee}_\perp\right)^2}\,.
\end{equation}
Using Eq.~(\ref{pheno-conductivity2}), an identification of the e-e
contribution is possible in a perturbative regime where $1/\tau^{ee}_\perp,1/\tau^{\rm dis}_\perp\ll\omega_0$, leading to the following formula:
\begin{equation}\label{pheno-conductivity}
\sigma_\perp =  \frac{n}{4m_\ast\omega_0^2}
\left(\frac{1}{\tau^{\rm
dis}_\perp}+\frac{1}{\tau^{ee}_\perp}\right)\,.
\end{equation}
\begin{figure}[b]
\vskip -0.3 in
\includegraphics[width=3.5in]{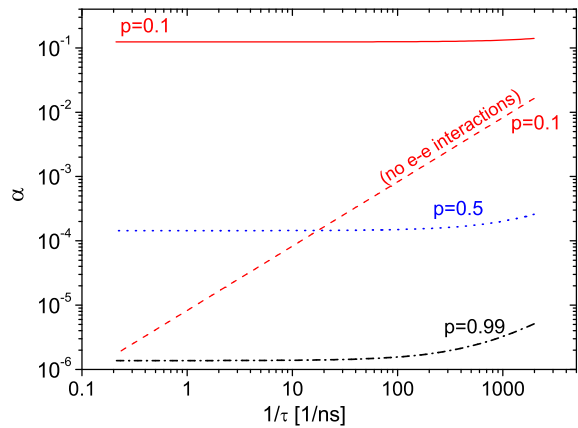}
\caption{(Color online) The Gilbert damping $\alpha$ as a function
of the disorder scattering rate $1/\tau$. Red (solid) line shows
the Gilbert damping for polarization $p=0.1$ in the presence of
the e-e and disorder scattering, while dashed line does not
include the e-e scattering. Blue (dotted) and black (dash-dotted)
lines show Gilbert damping for $p=0.5$ and $p=0.99$, respectively.
We took $q=0.1k_F$, $T=54K$, $\omega_0 =
E_F[(1+p)^{2/3}-(1-p)^{2/3}]$, $M_0 =\gamma pn/2$, $m_\ast= m_e$,
$n=1.4\times 10^{21}cm^{-3}$, $r_s=5$, $a_\ast =2a_0$}
\end{figure}
Comparison with Eq.~(\ref{force-force}) enables us to immediately
identify the microscopic expressions for the two scattering rates.
For the disorder contribution, we recover what we already knew,
i.e., Eq.~(\ref{tauperp}). For the e-e interaction contribution,
we obtain
\begin{eqnarray}\label{gammaperp}
\frac{1}{\tau^{ee}_\perp} =-\frac{4}{nm_\ast\cal V}\lim_{\omega\to
0} \frac{\Im m\langle \langle\sum_i \hat S_{ia}\hat
F^C_{ia};\sum_i \hat S_{ia}\hat
F^C_{ia}\rangle\rangle_{\omega}}{\omega}\,,
\end{eqnarray}
where $F^C$ is just the Coulomb force, and the force-force
correlation function is evaluated in the absence of disorder. The
correlation function in Eq.~(\ref{gammaperp}) is proportional to
the function $F_{+-}(\omega)$ which appeared in
Ref.~\onlinecite{Qian02} [Eqs. (18) and (19)] in a direct
calculation of the transverse spin susceptibility.   Making use of
the analytic result for $\Im m F_{+-}(\omega)$ presented in Eq.
(21) of that paper  we obtain
\begin{equation}\label{gammaperp2}
\frac{1}{\tau^{ee}_\perp} =\Gamma(p)\frac{8\alpha_0}{27}\frac{T^2
r_s^4 m_\ast a_\ast^2k_B^2}{(1+p)^{1/3}}\,,
\end{equation}
where $T$ is the temperature, $p=(n_{\uparrow}-n_{\uparrow})/n$ is
the degree of spin polarization, $a_\ast$ is the effective Bohr
radius, $r_s$ is the dimensionless Wigner-Seitz radius,
$\alpha_0=(4/9\pi)^{1/3}$ and $\Gamma(p)$  --  a dimensionless
function of the polarization $p$ --  is defined by Eq.~(23) of
Ref.~\onlinecite{Qian02}.  This result is valid to second order in
the Coulomb interaction. Collecting our results, we finally obtain
a full expression for the $q^2$ Gilbert damping parameter:
\begin{equation}\label{Gilbert2}
\alpha = \frac{\gamma n q^2}{4m_\ast M_0} \frac{1/\tau^{\rm
dis}_\perp+1/\tau^{ee}_\perp}{\omega_0^2+\left(1/\tau^{\rm
dis}_\perp+1/\tau^{ee}_\perp\right)^2}\,.
\end{equation}
One of the salient features of Eq.~(\ref{Gilbert2}) is that it
scales as the total scattering \textit{rate} in the weak disorder and e-e
interactions limit, while it scales as the scattering \textit{time} in the opposite limit.
The approximate formula for the Gilbert damping in the more  interesting weak-scattering/strong-ferromagnet  regime is
\begin{equation}\label{Gilbert_weak}
\alpha = \frac{\gamma
nq^2}{4m_\ast\omega_0^2M_0}\left(\frac{1}{\tau^{\rm dis}_\perp}+
\frac{1}{\tau^{ee}_\perp}\right)\,,
\end{equation}
while in the opposite limit, i.e. for $\omega_0 \ll1/\tau^{\rm
dis}_\perp,1/\tau^{ee}_\perp$:
\begin{equation}\label{Gilbert_strong}
\alpha =\frac{\gamma nq^2}{4m_\ast M_0}\left(\frac{1}{\tau^{\rm
dis}_\perp}+\frac{1}{\tau^{ee}_\perp}\right)^{-1}.
\end{equation}
Our Eq.~(\ref{Gilbert2}) agrees with the result of Singh and
Te{\v{s}}anovi{\'c}\cite{Tesanovic89} on the spin-wave linewidth
as a function of the disorder strength and $\omega_0$. However,
Eq.~(\ref{Gilbert2}) also describes the influence of e-e
correlations on the Gilbert damping. A comparison of the
scattering rates originating from disorder and e-e interactions
shows that the latter is important and can be comparable or even
greater than the disorder contribution for high-mobility and/or
low density 3D metallic samples. Fig.~1 shows the behavior of the
Gilbert damping as a function of the disorder scattering rate. One
can see
 that the e-e scattering  strongly enhances the Gilbert damping for small
 polarizations/weak ferromagnets, see the red (solid) line.
 This stems from the fact that
$1/\tau^{\rm dis}_\perp$ is proportional to $1/\tau$ and
independent of polarization for small polarizations, while
$1/\tau^{ee}_\perp$ is enhanced by a large prefactor
$\Gamma(p)=2\lambda/(1-\lambda^2)+(1/2)\ln[(1+\lambda)/(1-\lambda)]$,
where $\lambda = (1-p)^{1/3}/(1+p)^{1/3}$. On the other hand, for
strong polarizations (dotted and dash-dotted lines in Fig.~1), the
disorder dominates in a broad range of $1/\tau$ and the
inhomogenous contribution to the Gilbert damping is rather small.
Finally, we note that our calculation of the e-e interaction
contribution to the Gilbert damping is valid under the assumption
of $\hbar\omega \ll k_BT$ (which is certainly the case if
$\omega=0$). More generally, as follows from Eqs.~(21) and (22) of
Ref.~\onlinecite{Qian02}, a finite frequency $\omega$ can be
included through the replacement $(2 \pi k_BT)^2 \to  (2 \pi
k_BT)^2+(\hbar\omega)^2$ in Eq.~(\ref{gammaperp2}).   Thus
$1/\tau_\perp^{ee}$ is proportional to the scattering rate of
quasiparticles near the Fermi level, and our damping constant in
the clean limit becomes qualitatively similar to the damping
parameter obtained by Mineev\cite{Mineev04} for $\omega$
corresponding to the spin-wave resonance condition in some
external magnetic field (which in practice is much smaller than
the ferromagnetic exchange splitting $\omega_0$).

 {\em Summary} --
We have presented a unified theory of the Gilbert damping in
itinerant electron ferromagnets at the order $q^2$, including e-e
interactions and disorder on equal footing. For the inhomogeneous
dynamics ($q \neq 0$),
 these processes add to a $q=0$ damping contribution that is governed by magnetic disorder and/or spin-orbit interactions.  We have shown that the calculation of the
 Gilbert damping can be formulated in the language of  the spin conductivity, which
takes  an intuitive Matthiessen form with the disorder and
interaction contributions being simply additive. It is still a
common practice, e.g., in the micromagnetic calculations of
spin-wave dispersions and linewidths, to use a Gilbert damping
parameter independent of $q$. However, such calculations are often
at odds with experiments
 on the quantitative side, particularly where the linewidth is concerned.\cite{Krivorotov07}
  We suggest that the inclusion of the $q^2$ damping (as well as the associated magnetic noise) may help in reconciling theoretical calculations with
  experiments.

{\it Acknowledgements} -- This work was supported in part by NSF
Grants Nos.~DMR-0313681 and DMR-0705460 as well as Fordham
Research Grant. Y. T. thanks A. Brataas and G. E. W. Bauer for
useful discussions.

\end{document}